\documentclass[12pt]{article}
\usepackage{graphicx}
\usepackage{epstopdf}
\usepackage{amsfonts}
\usepackage{amssymb}
\usepackage[footnotesize]{caption2}
\usepackage{mathrsfs}
\usepackage{amsmath}

\usepackage{authblk}

\begin{document}
\renewcommand{\thefootnote}{\fnsymbol{footnote}}
\title{Maximally localized states and quantum corrections of black hole thermodynamics in the extreme case with an improved exponential GUP}
\author[1]{\small Ying-Jie Zhao \thanks{E-mail address: yj\_zhao@bjut.edu.cn}}

\affil[1]{\small Institute of Theoretical Physics, Beijing University of Technology,  Beijing 100124, China}

\date{}
\maketitle
\begin{abstract}

\setlength{\parindent}{0pt} \setlength{\parskip}{1.5ex plus 0.5ex
minus 0.2ex} 
We have introduced an improved exponential GUP, derived the maximally
localized states, calculated quantum corrections to the thermodynamic quantities of the Schwardzschild
black hole in our previous work. In this paper we continue to investigate how the maximally
localized states and thermodynamic quantities such as Hawking temperature, the entropy, the heat capacity, the evaporation rate, and the decay time change in the extreme case that the integer $n$ in our GUP rises to infinity.
\vskip 10pt
\noindent
{\bf PACS Number(s)}: 02.40.Gh, 04.70.Dy, 03.65.Sq
\vskip 5pt
\noindent

\end{abstract}

\thispagestyle{empty}

\newpage

\setcounter{page}{1}

\section{Introduction}
Many approaches to unify the quantum mechanics and general relativity, including string theory \cite{s1}\cite{s2}\cite{s3}\cite{s4}, loop quantum gravity \cite{s5}, and quantum geometry \cite{s6} have attracted much attention in recent years. Almost all these proposals, together with some experiments \cite{s7}, support a minimal length of the order of Planck scale and a modification of the Heisenberg uncertainty principle (HUP) to the so-called generalized uncertainty principle (GUP)\cite{s8}\cite{s9}\cite{s10}\cite{s11}\cite{s12}.
In the framework of GUP, contrary to the HUP's one, a lower bound of the measurable length of the order of the Planck scale $10^{-35}m$ naturally appears in the spacetime\cite{s10}\cite{s11}\cite{s12}\cite{s13}.

Generalized uncertainty principle, being one of the approaches in adding quantum effects on gravity systems, have received much attention and several achievements have been made \cite{s15}\cite{s16}\cite{s17}\cite{s18}. For instance, in the framework of GUP, the existence of a minimal length prohibits the wholly radiation of a black hole and there finally remains a remnant that the information loss paradox may be avoided.

Various typical forms of GUP deforming the position-momentum commutation relations  have been mentioned by researchers. One is the quadratic form GUP \cite{s8} proposed by KMM in which the commutators of position and momentum operators contain a additional quadratic term in momentum. Another is the exponential form proposed by K. Nouicer \cite{s19} in which GUP deforms the commutation relation by an exponential function of the square of momentum operator. In our previous work \cite{s20}\cite{yys}, we have introduced an improved exponential form GUP and obtain the cosmological constant that coincides exactly with the experimental value provided
by the Planck 2013 results \cite{Planck} by choosing an appropriate index $n$ in our GUP while considering the UV/IR mixing effect. Moreover, we have investigated the maximally localized states, the corresponding quasi-position wavefunctions, and the scalar product
of these wavefunctions, derived the corrected thermodynamic quantities of the Schwardzschild black hole with or without considering the UV/IR mixing effect. We have analyzed these results in different cases of index $n$ and made some interesting conclusions.

However, in our previous paper only the situation of finite index $n$ has been studied in detail. How will the situations of maximally localized states and quantum corrections of black hole thermodynamics in the framework of GUP change as $n$ in our GUP rises to infinity? The paper answering this question is organized as follows. In Section $2$ we briefly review our improved exponential GUP researched in detail in refs.\cite{s20}\cite{yys}. In section $3$, we discuss the representation of operators and the maximally localized states. In section $4$, we then focus on the corrected Hawking temperature, entropy, and heat capacity of the Schwarzschild black hole; we also pay attention to the Hawking evaporation process and calculate the
corrections to the evaporation rate and the decay time without and with the consideration of the UV/IR
mixing effect. In the above two sections all the results are derived in the extreme case, $n \rightarrow \infty$.  At the end, we make a short conclusion in section 5.

\section{The improved exponential GUP}
In this section we start with the improved exponential GUP introduced in paper \cite{yys}.
We propose our  improved exponential GUP as follows,
\begin{eqnarray}
[ {\hat{X}, \hat{P}} ] = i\hbar \exp \left( \frac{{{\alpha ^{2n}}\ell_{\rm Pl}^{2n}}}{{{\hbar ^{2n}}}}{{\hat{P}^{2n}}} \right), \label{XandP}
\end{eqnarray}
where  $\alpha$ is a dimensionless parameter with the order of unity that describes the strength of gravitational effects,
$\ell_{\rm Pl}$ is the Planck length,
and $n$ is a positive integer.

In the momentum space, the position and momentum operators of our new GUP form can be represented as
\begin{eqnarray}
\hat{X} \psi (p) &=& i\hbar \exp \left( {\frac{{{\alpha ^{2n}}\ell_{\rm Pl}^{2n}}}{{{\hbar ^{2n}}}}{p^{2n}}} \right){\partial _p}\psi (p),
\label{Xp}\\
\hat{P} \psi (p) &=& p\,\psi (p), \label{Pp}
\end{eqnarray}
and the symmetric condition~\cite{s8}
\begin{eqnarray}
\left( {\left\langle \phi  \right|\hat{X}} \right)\left| \psi  \right\rangle  = \left\langle \phi  \right|\left( {\hat{X}\left| \psi  \right\rangle } \right), \qquad
\left( {\left\langle \phi  \right|\hat{P}} \right)\left| \psi  \right\rangle  = \left\langle \phi  \right|\left( {\hat{P}\left| \psi  \right\rangle } \right),
\end{eqnarray}
gives rise to the following scalar product of wavefunctions and the orthogonality and completeness of eigenstates,
\begin{eqnarray}
\left\langle \phi | \psi \right\rangle  &=& \int_{ - \infty }^{ + \infty } {dp \,\exp \left( { - \frac{{{\alpha ^{2n}}\ell_{\rm Pl}^{2n}}}{{{\hbar ^{2n}}}}{p^{2n}}} \right){\phi ^*}(p)\,\psi (p)}, \\
\left\langle p | p' \right\rangle  &=& \exp \left( {\frac{{{\alpha ^{2n}}\ell_{\rm Pl}^{2n}}}{{{\hbar ^{2n}}}}{p^{2n}}} \right)\delta \left( {p - p'} \right), \\
1 &=& \int_{ - \infty }^{ + \infty } { dp\, \exp \left( { - \frac{{{\alpha ^{2n}}\ell_{\rm Pl}^{2n}}}{{{\hbar ^{2n}}}}{p^{2n}}} \right) {| p \rangle } {\langle  p |}},
\end{eqnarray}
where $|p \rangle$ and $|p' \rangle$ mean momentum eigenstates and $\psi (p) \equiv \langle p | \psi \rangle$ stands for  a wavefunction in the momentum space.
From the improved exponential GUP we receive the uncertainty relation,
\begin{eqnarray}
\Delta  {\hat X} \Delta  {\hat P} \geqslant \frac{\hbar}{2} {\bigg \langle} \exp \left( \frac{{{\alpha ^{2n}}\ell_{\rm Pl}^{2n}}}{{{\hbar ^{2n}}}}{{\hat{P}^{2n}}}\right) {\bigg\rangle},
\end{eqnarray}
Using the relations $\langle{\hat P}^2\rangle = \langle{\hat P}\rangle^2 + (\Delta P)^2$ and $\langle{\hat P}^{2n}\rangle \geqslant \langle{\hat P}\rangle^{2n}$,
the minimal length and its corresponding momentum measurement precision are straightly expressed by choose $\langle \hat{P} \rangle =0 $,
\begin{eqnarray}
(\Delta X)_0 = \frac{\alpha \ell_{\rm Pl}}{2} (2 n e)^{\frac{1}{2n}}, \quad (\Delta P)_C =\left( \frac{1}{2n}\right)^{\frac{1}{2n}}
\frac{\hbar}{\alpha\ell_{ \rm Pl} }.
\end{eqnarray}

\section{Representation and maximally localized states in extreme case $n \rightarrow \infty$ }
In KMM's opinion \cite{s8}, the existence of the minimal length fuzzes the concept of eigenvalues of the position operator and the formal eigenfunction cannot be worked out by directly solving the eigenvalue equation. Factually, the maximally localized states are imperative to recover the information on position by calculating the average values of the position operator rather than the ordinary position eigenvalues via position eigenstates. The maximally localized states in our improved exponential GUP, unfortunately, cannot simply worked our in KMM's approach for GUP's complex from. S. Detournay applies a constrained variational principle to find out maximally localized states satisfying an Euler-Lagrange equation in the momentum space written as a piecewise-defined function (See details in \cite{yys}\cite{Detournay}),

\begin{scriptsize}
\begin{eqnarray}
\Psi_\xi(p) = - \sqrt{\frac{\alpha \ell_{\rm Pl}}{\hbar\Gamma(\frac{2n+1}{2n})}}\exp\left[{\frac{i \xi \Gamma(\frac{2n+1}{2n})}{\alpha \ell_{\rm Pl}}} + \frac{i\xi p }{2n \hbar}E_{\frac{2n-1}{2n}}\left( \frac{{{\alpha ^{2n}}\ell_{\rm Pl}^{2n}}}{{{\hbar ^{2n}}}}{{p^{2n}}} \right) \right] \sin\left[\frac{\pi \alpha \ell_{\rm Pl}p }{4 n \hbar \Gamma(\frac{2n+1}{2n})}E_{\frac{2n-1}{2n}}\left( \frac{{{\alpha ^{2n}}\ell_{\rm Pl}^{2n}}}{{{\hbar ^{2n}}}}{{p^{2n}}} \right)\right]
\end{eqnarray}
\end{scriptsize}
for $p<0$, and
\begin{scriptsize}
\begin{eqnarray}
\Psi_\xi(p) =   \sqrt{\frac{\alpha \ell_{\rm Pl}}{\hbar\Gamma(\frac{2n+1}{2n})}}\exp\left[-{\frac{i \xi \Gamma(\frac{2n+1}{2n})}{\alpha \ell_{\rm Pl}}} + \frac{i\xi p }{2n \hbar}E_{\frac{2n-1}{2n}}\left( \frac{{{\alpha ^{2n}}\ell_{\rm Pl}^{2n}}}{{{\hbar ^{2n}}}}{{p^{2n}}} \right) \right] \sin\left[\frac{\pi \alpha \ell_{\rm Pl}p }{4 n \hbar \Gamma(\frac{2n+1}{2n})}E_{\frac{2n-1}{2n}}\left( \frac{{{\alpha ^{2n}}\ell_{\rm Pl}^{2n}}}{{{\hbar ^{2n}}}}{{p^{2n}}} \right)\right]
\end{eqnarray}
\end{scriptsize}
for $p\geqslant 0$. The position $\xi$ denotes the normalized average value of the position operator $\langle \Psi_\xi | \hat{X}| \Psi_\xi  \rangle$. Here we discover the probability density $|\Psi_\xi(p)|^2$ at the points $|p|>\frac{\hbar}{\alpha \ell_{\rm Pl}}$ will markedly plunge when $n$ grows, and at the moment $n$ goes to infinity it will vanish. The fact that the zero probability density  outside the range indicates the momentum should be given a natural limitation in this extreme case
\begin{eqnarray}
|p|\leqslant\frac{\hbar}{\alpha \ell_{\rm Pl}}.
\end{eqnarray}
Besides, it ensure the position and momentum operators acting on wavefunctions in momentum space have a physical interpretation as $n \rightarrow \infty$.
We notice that the momentum interval is the order of the Plack momentum. In this case, GUP seems to reduce to the normal form,
\begin{eqnarray}
[ {\hat{X}, \hat{P}} ] = i\hbar,
\end{eqnarray}
and so do the representations of the position and momentum operators and the scalar product of wavefunctions and the orthogonality and completeness of eigenstates take the following forms,
\begin{eqnarray}
\hat{X} \psi (p) = i\hbar \psi (p),   \quad \quad
 \hat{P} \psi (p) = p\,\psi (p),
\end{eqnarray}
\begin{eqnarray}
\left\langle \phi | \psi \right\rangle  = \int_{ - \frac{\hbar}{\alpha \ell_{\rm Pl}} }^{ \frac{\hbar}{\alpha \ell_{\rm Pl}} } {dp \,{\phi ^*}(p)\, \psi (p)}, \quad \quad
\left\langle p | p' \right\rangle  = \delta \left( {p - p'} \right), \quad \quad
1 = \int_{ - \frac{\hbar}{\alpha \ell_{\rm Pl}}}^{  \frac{\hbar}{\alpha \ell_{\rm Pl}} } { dp\,  {| p \rangle } {\langle  p |}}.
\end{eqnarray}
However, it is noted that the above results are only valid in the momentum interval, although they look like their counterparts in the ordinary quantum mechanics. The

In S. Detournay's way \cite{Detournay}, we can ensure from eqs.(20)(25) in ref.\cite{yys} that $z(p)$ is a linear function of momentum in the case $n$ inclining to infinity,
\begin{eqnarray}
\lim_{n\rightarrow\infty}{z(p)} = \frac{p}{\hbar},
\end{eqnarray}
and therefore obtain from eq.(28)(29) in ref.\cite{yys} the maximally localized states,
\begin{eqnarray}
\Psi _\xi (p) = \sqrt{ \frac{\alpha \ell_{\rm Pl}}{\hbar}} \exp  \left( - \frac{i \xi p }{\hbar}\right)
\cos \left(   \frac{\pi \alpha \ell_{\rm Pl} }{2\hbar}p\right),
\end{eqnarray}
and the minimal length (eq.(24) in ref.\cite{yys}),
\begin{eqnarray}
\left(\Delta X\right)_{min} {\bigg |}_{b = 0} = \frac{\pi \alpha \ell_{\rm Pl} }{2}.
\end{eqnarray}

Moreover, we list some properties of the maximally localizes states in the extreme case $n \rightarrow \infty$.
At first, eq.(31) in ref.\cite{yys} reduces to
\begin{eqnarray}
\langle\Psi _{\xi'} | \Psi _{\xi}\rangle  =
\frac{{\pi ^2}{\alpha ^3}{\ell_{\rm Pl}^3}} {{\pi ^2}{\alpha ^2}\ell_{\rm Pl}^2\left( {\xi  - \xi '} \right) - {{\left( {\xi  - \xi '} \right)}^3}}\sin \left({\frac{ {\xi  - \xi '}}{{\alpha \ell_{\rm Pl}}}}\right),
\end{eqnarray}
which shows that the maximal localization states still remain non-orthogonal. Then, the transformation of a wavwfuntion from the momentum space into the quasi-position space, eq.(32) in ref.\cite{yys} , is simplified to be
momentum space to the quasi-position space,
\begin{eqnarray}
\psi(\xi) &\equiv& \langle \Psi _\xi  | \psi \rangle    \nonumber \\
&=&  \sqrt{ \frac{\alpha \ell_{\rm Pl}}{\hbar}}   \int_{  - \frac{\hbar}{\alpha \ell_{\rm Pl}} }^{ \frac{\hbar}{\alpha \ell_{\rm Pl}}} dp \exp  \left(  \frac{i \xi p }{\hbar}\right)
\cos \left(   \frac{\pi \alpha \ell_{\rm Pl} }{2\hbar}p\right) \psi(p),
\end{eqnarray}
and correspondingly the inverse transformation, i.e. eq.(34) in ref.\cite{yys}  reads as
\begin{eqnarray}
\psi(p)
= \frac{1}{2\pi\sqrt{  \alpha \ell_{\rm Pl}\hbar} }\sec \left(   \frac{\pi \alpha \ell_{\rm Pl} }{2\hbar}p\right) \int_{  - \infty}^{ +\infty} d\xi\exp  \left( - \frac{i \xi p }{\hbar}\right)
 \psi(\xi),
\end{eqnarray}
The last property listed here is the simplified form of the scalar product of quasi-position wavefunction (eq.(35) in ref.\cite{yys})  in the limit
\begin{small}
\begin{eqnarray}
 {\langle \phi  | \psi \rangle} &=& \int_{ - \frac{\hbar}{\alpha \ell_{\rm Pl} }}^{  \frac{\hbar}{\alpha \ell_{\rm Pl} } }{dp}
 \, {\phi ^*}(p)\,\psi(p) \nonumber \\
&=& \frac{1}{4{\pi}^2 \alpha \ell_{\rm Pl} \hbar} \int_{ - \frac{\hbar}{\alpha \ell_{\rm Pl} }}^{  \frac{\hbar}{\alpha \ell_{\rm Pl} } } {dp} \int_{ - \infty }^{ + \infty } {d\xi} \int_{ - \infty }^{ + \infty } {d\xi'} \,\exp\left[\frac{i(\xi'-\xi)p}{\hbar} \right] \sec^2\left( \frac{\pi \alpha \ell_{\rm Pl} }{2\hbar}p\right)  {\phi ^*}(\xi')\,\psi(\xi).
\end{eqnarray}
\end{small}

\section{Black hole thermodynamics in extreme case $n \rightarrow \infty$ }
At the beginning we write the metric of a four-dimensional Schwarzschild black hole,
\begin{eqnarray}
ds^2 = - \left( 1- \frac{2GM}{r}\right)dt^2 +\left( 1- \frac{2GM}{r}\right)^{-1}dr^2+ r^2 d{\Omega}^2,
\end{eqnarray}
where $M$ denotes the black hole mass. Considering the near-horizon geometry the accuracy of position measurement should be of the scale of the black hole, $\Delta X \simeq r_h = 2GM$, hence the minimal mass has been given
\begin{eqnarray}
M_0 = \frac{\alpha M_{\rm Pl}}{4} (2ne)^{\frac{1}{2n}}.
\end{eqnarray}
And the minimal mass (Black hole remnant) is $M_0 = \frac{\alpha M_{\rm Pl}}{4}$ as $n$ tends to infinity.
With the help of the Lambert W function $W(x)$ \cite{s21}, the corrected temperature in ref.\cite{yys} is shown as follows,
\begin{eqnarray}
T = \frac{1}{8 \pi M G} \exp \left\{-\frac{1}{2n} W\left(-\frac{1}{e}\left(\frac{M_0}{M}\right)^{2n}\right)  \right\},
\end{eqnarray}
from which we obtain the temperature and and the maximal temperature in the framework of GUP, respectively,
\begin{eqnarray}
 T = \frac{1}{8 \pi M \ell_{\rm Pl}^2}, \quad T_{max} = \frac{T_{\rm Pl}}{2\pi \alpha}, \label{TT}
\end{eqnarray}
where $T_{max}$ is the highest temperature of all the black hole remnants. Moreover, using
\begin{eqnarray}
S =  \frac{1}{4\ell_{\rm Pl}^2} \int_{A_0}^{A}  \exp \left\{ \frac{1}{2n} W\left(-\frac{1}{e}\left(\frac{A_0}{A}\right)^{2n}\right)  \right\} dA ,
\end{eqnarray}
in ref.\cite{yys}, we derive the entropy of the black hole,
\begin{eqnarray}
S = \frac{A- {A_0}}{4\ell_{\rm Pl}^2} =  \frac{\pi \alpha^2}{4}\left[ \left( \frac{M}{M_0}\right)^2 -1 \right],
\end{eqnarray}
and from
\begin{eqnarray}
C = -8 \pi M^2 {\ell_{\rm Pl}^2}\left\{ 1+ W\left(-\frac{1}{e}\left(\frac{M_0}{M}\right)^{2n}\right)  \right\}  \exp \left\{ \frac{1}{2n} W\left(-\frac{1}{e}\left(\frac{M_0}{M}\right)^{2n}\right)  \right\}  ,
\end{eqnarray}
in ref.\cite{yys}, the heat capacity is
\begin{eqnarray}
C = -8\pi M^2 \ell_{\rm Pl}^2,
\end{eqnarray}
In contrast to the situation with a finite $n$, the heat capacity of the black hole does not vanish, whose absolute value equals
$\pi \alpha^2/2$ due to the non-vanishing remnant mass $M_0 = \alpha M_{Pl}/4$.

Now we turn to the discussion of the evaporation rate and the decay time in the GUP framework. Setting $z\equiv P/T$ in eq.(50) in ref.\cite{yys},
we can write the evaporation rate without the UV/IR mixing effect as
\begin{eqnarray}
\frac{dM}{dt} = -\frac{4 G^2 M^2 T^4}{\pi} \int_{0}^{\infty}e^{-3(\frac{\alpha T z}{T_{\rm Pl}})^{2n}}\frac{z^3 dz}{e^{z} -1},
\end{eqnarray}
When $n$ climbs to infinity, the above integration vanishes unless the upper limit is not greater than $\frac{T_{Pl}}{\alpha T}$. Again using the second equation of eq.(\ref{TT}), we can perform the integration
\begin{eqnarray}
\frac{dM}{dt} &=& -\frac{1}{1024 \pi^5 M^2 G^2} \int_{0}^{\frac{8\pi M}{\alpha T_{\rm Pl}}}\frac{z^3 dz}{e^{z} -1}  \nonumber \\
&=& -\frac{1}{1024 \pi^5 M^2 G^2} \left[ -\frac{ \pi^4}{15} + \left(\frac{8 \pi M}{\alpha T_{\rm Pl}}\right)^4 -\left(\frac{8 \pi M}{\alpha T_{\rm Pl}}\right)^3 \ln\left(e^\frac{8 \pi M}{\alpha T_{\rm Pl}}-1\right) \right.  \nonumber \\
& &\left.+ 3\left(\frac{8 \pi M}{\alpha T_{\rm Pl}}\right)^2 Li_2
\left(- e^\frac{8 \pi M}{\alpha T_{\rm Pl}}\right)+ 6\left(\frac{8 \pi M}{\alpha T_{\rm Pl}}\right)  Li_3
\left(- e^\frac{8 \pi M}{\alpha T_{\rm Pl}}\right)+ 6 Li_4
\left(- e^\frac{8 \pi M}{\alpha T_{\rm Pl}}\right) \right],
\end{eqnarray}
where $Li_n (x)\equiv \sum_{k=1}^{\infty} \frac{x^k}{k^n}$ is called the polylogarithm function of order $n$.
Correspondingly, we can calculate the decay time by numerical method and list the results in Table 1.

\begin{table}[!htbp]
\small
\centering
\begin{tabular}{|c|*{6}{|c}}
\hline

M & $1$ & $2$ & $3$ & $4$ & $5$  \\ \hline \hline
Hawking & $16085$ & $128680$ & $434294$ & $1.02944\times10^6$ & $2.01062\times10^6 $ \\ \hline
$\mathrm{GUP}$ & $15868.8$ & $ 128463$ & $  434078 $ & $ 1.02922\times10^{6}$ & $ 2.0104\times10^{6}$ \\ \hline
 \end{tabular}
\caption{Hawking time and $\mathrm{GUP}$-corrected decay times with the black hole mass $M=1, 2, \cdots, 5$ (in Planck units) and $\alpha=1$, $n\rightarrow\infty$.}
\end{table}

Finally, we analyze the evaporation rate and the decay time with the consideration of the UV/IR mixing effect.
By setting $z \equiv P/T$ and taking the limit $n \rightarrow \infty$ in eq.(52) in ref.\cite{yys}
\begin{eqnarray}
\frac{dM}{dt} = -\frac{4 G^2 M^2 }{\pi} \int_{(\frac{1}{2n})^{1/(2n)}\frac{1}{\alpha \ell_{\rm Pl}}}^{\infty} \lim_{n \rightarrow \infty} e^{-3 {\alpha}^{2n} \ell_{\rm Pl}^{2n}P^{2n}} \frac{P^3 dP}{e^{P/T} -1},
\end{eqnarray}
 and further using the fist equation of eq.(\ref{TT}),
we obtain the expression of the evaporation rate,
\begin{eqnarray}
\frac{dM}{dt} = -\frac{1}{1024 \pi^5 G^2 M^2} \int_{\frac{8 \pi M}{\alpha T_{\rm Pl}}}^{\infty} \lim_{n \rightarrow \infty} e^{-3(\frac{\alpha T_{\rm Pl} z}{8 \pi M})^{2n}} \frac{z^3 dz}{e^{z} -1}.
\end{eqnarray}
Because  the range of integration is larger than $\frac{8 \pi M }{\alpha T_{\rm Pl}}$, which implies that only the trans-Planckian modes \cite{s22} ($P > P_C$) have contributions to the energy density, we have
\begin{eqnarray}
\lim_{n \rightarrow \infty} e^{-3(\frac{\alpha T_{\rm Pl} z}{8 \pi M})^{2n}} =0,
\end{eqnarray}
and hence deduce that the evaporation rate is vanishing. In other words, the decay time slides grows to infinity, i.e. the black hole has no radiation in this extreme case.

\section{Conclusion}

In this paper we first review our improved exponential GUP in our previous work, and analyze in the ultimate case how the representation of operators and the properties of the maximally localized states alter. As the integer $n\rightarrow \infty$ some properties of the states such as the non-orthogonality, the corresponding quasi-position wavefunctions, and the scalar product of these wavefunctions have transformed. Especially, the existence of the momentum range can be seen as the first new physical feature.

Next, we investigate thermodynamics of the Schwardzschild black hole and find that in this extreme case the heat capacity of the black hole remnant as been limited not to vanish, which means the second new physical feature.

At last, we also have calculated the evaporation rates and decay times without and with UV/IR mixing effect in the case $n\rightarrow \infty$, which has made of great differences. In the former situation, the black hole evaporation rate is decided by its mass and the time is approximately to ordinary one, but in the latter situation, the black hole stops emission and remains its original mass forever. The third new physical feature has emerged.


\begin{thebibliography}{99}

\bibitem{s1}G. Veneziano, {\em A stringy nature needs just two constants}, Europhys. Lett. \textbf{2}, 199 (1986).
\bibitem{s2}E. Witten, {\em Reflections on the fate of spacetime}, Phys. Today \textbf{49} (4), 24 (1996).
\bibitem{s3}D. Atami, M. Ciafaloni, and G. Veneziano, {\em Superstring collisions at planckian energies}, Phys. Lett. B \textbf{197}, 81 (1987); \\
D. Atami, M. Ciafaloni, and G. Veneziano, {\em Can spacetime be probed below the string size?} Phys. Lett. B \textbf{216}, 41 (1989);\\
D. Atami, M. Ciafaloni, and G. Veneziano, {\em Higher-order gravitational deflection and soft bremsstrahlung in planckian energy superstring collisions}, Nucl. Phys. B \textbf{347}, 550 (1990).
\bibitem{s4}K. Konishi, G.Paffuti, and P. Provero, {\em Minimum physical length and the generalized uncertainty principle in string theory}, Phys. Lett. B \textbf{234}, 276 (1990).
\bibitem{s5}L.J. Garay, {\em Quantum gravity and minimum length}, Int. J. Mod. Phys. A \textbf{10}, 145 (1995) [arXiv:gr-qc/9403008].
\bibitem{s6}S. Capozziello, G. Lambiase, and G. Scarpetta, {\em The generalized uncertainty principle from quantum geometry}, Int. J. Theor. Phys. \textbf{39}, 15 (2000).
\bibitem{s7}F. Scardigli, {\em Generalized uncertainty principle in quantum gravity from micro-black hole gedanken experiment}, Phys. Lett. B \textbf{452}, 39 (1999) [arXiv:hep-th/9904025].
\bibitem{s8}A. Kempf, G. Mangano, and R.B. Mann, {\em Hilbert space representation of the minimal length uncertainty relation}, Phys. Rev. D \textbf{52}, 1108 (1995) [arXiv:hep-th/9412167].
\bibitem{s9}A. Kempf and G. Mangano, {\em Minimal length uncertainty relation and ultraviolet regularization}, Phys. Rev. D \textbf{55},  7909 (1997) [arXiv:hep-th/9612084].
\bibitem{s10}M. Maggiore, {\em A generalized uncertainty principle in quantum gravity}, Phys. Lett. B \textbf{304}, 65 (1993) [arXiv:hep-th/9301067].
\bibitem{s11}M. Maggiore, {\em The algebraic structure of the generalized uncertainty principle}, Phys. Lett. B \textbf{319}, 83 (1993)  [arXiv:hep-th/9309034].
\bibitem{s12}M. Maggiore, {\em Quantum groups, gravity, and the generalized uncertainty principle}, Phys. Lett. B \textbf{49}, 5182 (1994) [arXiv:hep-th/9305163].
\bibitem{s13}S. Hossenfelder, {\em Can we measure structures to a precision better than the Planck length?} Class. Quantum Grav. \textbf{29}, 115011 (2012) [arXiv:1205.3636 [gr-qc]].
\bibitem{s15}P. Pedram, {\em New approach to nonperturbative quantum mechanics with minimal length uncertainty}, Phys. Rev. D \textbf{85}, 024016 (2012) [arXiv:1112.2327 [hep-th]]; \\
P. Pedram, {\em A higher order GUP with minimal length uncertainty and maximal momentum}, Phys. Lett. B \textbf{714}, 317 (2012)  [arXiv:1110.2999 [hep-th]]; \\
P. Pedram, {\em A higher order GUP with minimal length uncertainty and maximal momentum II: Applications}, Phys. Lett. B \textbf{718}, 638 (2012) [arXiv:1210.5334 [hep-th]].
\bibitem{s16}S. Das and E.C. Vagenas, {\em Universality of quantum gravity corrections}, Phys. Rev. Lett. \textbf{101}, 221301 (2008)  [arXiv:0810.5333 [hep-th]]; \\
S. Das and E.C. Vagenas, {\em Phenomenological implications of the generalized uncertainty principle}, Can. J. Phys. \textbf{87}, 233 (2009) [arXiv:0901.1768 [hep-th]]; \\
A.F. Ali, S. Das, and E.C. Vagenas, {\em Discreteness of space from the generalized uncertainty principle}, Phys. Lett. B \textbf{678}, 497 (2009) [arXiv:0906.5396 [hep-th]];  \\
S. Das, E.C. Vagenas, and A.F. Ali, {\em Discreteness of space from GUP II: relativistic wave equations}, Phys. Lett. B \textbf{690}, 407 (2010) [arXiv:1005.3368 [hep-th]];  \\
A.F. Ali, S. Das, and E.C. Vagenas, {\em Proposal for testing quantum gravity in the lab}, Phys. Rev. D \textbf{84}, 044013 (2011) [arXiv:1107.3164 [hep-th]].
\bibitem{s17}R.J. Adler, P. Chen, and D.I. Santiago, {\em The generalized uncertainty principle and black hole remnants}, Gen. Rel. Grav. \textbf{33}, 2101 (2001) [arXiv:gr-qc/0106080]; \\
K. Nozari and A.S. Sefiedgar, {\em Comparison of approaches to quantum correction of black hole thermodynamics}, Phys. Lett. B \textbf{635}, 156 (2006) [arXiv:gr-qc/0601116];  \\
K. Nozari and A.S. Sefiedgar, {\em On the existence of the logarithmic correction term in black hole entropy-area relation}, Gen. Rel. Grav. \textbf{39}, 501 (2007) [arXiv:gr-qc/0606046];  \\
W. Kim, E.J. Son, and M. Yoon, {\em Thermodynamics of a black hole based on a generalized uncertainty principle}, JHEP \textbf{0801}, 035 (2008) [arXiv:0711.0786 [gr-qc]];   \\
K. Nozari and S.H. Mehdipour, {\em Hawking radiation as quantum tunneling from a noncommutative Schwarzschild black hole}, Class. Quantum Grav. \textbf{25}, 175015 (2008) [arXiv:0801.4074 [gr-qc]];  \\
K. Nozari and S.H. Mehdipour, {\em Quantum gravity and recovery of information in black hole evaporation}, Europhys. Lett. \textbf{84}, 20008 (2008) [arXiv:0804.4221 [gr-qc]];  \\
L. Xiang and X.Q. Wen, {\em A heuristic analysis of black hole thermodynamics with generalized uncertainty principle}, JHEP \textbf{0910}, 046 (2009) [arXiv:0901.0603 [gr-qc]];   \\
K. Nozari and S.H. Mehdipour, {\em Parikh-Wilczek tunneling from noncommutative higher dimensional black holes}, JEHP \textbf{0903}, 061 (2009) [arXiv:0902.1945 [hep-th]];   \\
R. Benerjee and S. Ghosh, {\em Generalised uncertainty principle, remnant mass and singularity problem in black hole thermodynamics}, Phys. Lett. B \textbf{688}, 224 (2010) [arXiv:1002.2302 [gr-qc]].
\bibitem{s18}
 S.W. Hawking, {\it Black hole explosions}, Nature {\bf 248}, 30  (1974);\\
 S.W. Hawking, {\it Particle creation by black holes}, Commun. Math. Phys. {\bf 43}, 199 (1975)
(Erratum-ibid. {\bf 46}, 206 (1976)).
\bibitem{s19}K. Nouicer, {\em Quantum-corrected black hole thermodynamics to all orders in the Planck length}, Phys. Lett. B \textbf{646}, 63 (2007) [arXiv:0704.1261 [gr-qc]].
\bibitem{s20}Y.-G. Miao and Y.-J. Zhao, {\em Interpretation of the cosmological constant problem within the framework of generalized uncertainty principle}, Int. J. Mod. Phys. D \textbf{23}, 1450062 (2014)  [arXiv:1312.4118 [hep-th]].
\bibitem{yys}Yan-Gang Miao, Ying-Jie Zhao, and Shao-Jun Zhang, {\em Maximally localized states and quantum corrections of black hole thermodynamics in the framework of a new generalized uncertainty principle}, AHEP Volume 2015 (2015), Article ID 627264 [arXiv:hep-th/1410.4115].
\bibitem{Planck}Planck Collaboration, {\em Planck 2013 Results. XVI. Cosmological Parameters}, arXiv:1303.5076 [astro-ph.CO].
\bibitem{Detournay}S. Detournay, C. Gabriel, and Ph. Spindel, {\em About maximally localized states in quantum mechanics}, Phys. Rev. D \textbf{66}, 125004 (2002) [arXiv:hep-th/0210128].

\bibitem{s21}R.M. Colrless, G.H. Gonnet, D.E. Hare, D.J. Jerey, and D. E. Knuth, {\em On the Lambert W function}, Adv. Comput. Math. \textbf{5}, 329 (1996).
\bibitem{s22}L.N. Chang, Z. Lewis, D. Minic, and T. Takeuchi, {\em On the minimal length uncertainty relation and the foundations of string theory}, Adv. High Energy Phys. \textbf{2011}, 493514 (2011) [arXiv:1106.0068 [hep-th]].



\end{thebibliography}
\end{document}